# Direct growth of hexagonal boron nitride on photonic chips for high-throughput characterization

Evgenii Glushkov[1*], Noah Mendelson[2], Andrey Chernev[1], Ritika Ritika[2], Martina Lihter[1], Reza R. Zamani[3], Jean Comtet[4], Vytautas Navikas[1], Igor Aharonovich[2,5], Aleksandra Radenovic[1*]

[1] Laboratory of Nanoscale Biology, Institute of Bioengineering, École Polytechnique Fédérale de Lausanne (EPFL), 1015 Lausanne, Switzerland
[2] School of Mathematical and Physical Sciences, University of Technology Sydney, Ultimo, NSW, 2007, Australia
[3] Centre Interdisciplinaire de Microscopie Électronique (CIME), École Polytechnique Fédérale de Lausanne (EPFL), 1015 Lausanne, Switzerland
[4] Laboratory of Soft Matter Science and Engineering, ESPCI Paris, PSL University, CNRS, Sorbonne Université, 75005 Paris, France
[5] ARC Center of Excellence for Transformative Meta-Optical Systems (TMOS), Faculty of Science, University of Technology Sydney, Australia
Corresponding authors: evgenii.glushkov@epfl.ch and aleksandra.radenovic@epfl.ch

## Abstract

Adapting optical microscopy methods for nanoscale characterization of defects in two-dimensional (2D) materials is a vital step for photonic on-chip devices. To increase the analysis throughput, waveguide-based on-chip imaging platforms have been recently developed. Their inherent disadvantage, however, is the necessity to transfer the 2D material from the growth substrate to the imaging chip which introduces contamination, potentially altering the characterization results. Here we present a unique approach to circumvent these shortfalls by directly growing a widely-used 2D material (hexagonal boron nitride, hBN) on silicon nitride chips, and optically characterizing the defects in the intact as-grown material. We compare the direct growth approach to the standard wet transfer method, and confirm the clear advantages of the direct growth. While demonstrated with hBN in the current work, the method is easily extendable to other 2D materials.

# Introduction

Characterizing defects in as-grown crystalline materials is a crucial step towards their further use in devices and applications. This is especially true for the rapidly developing field of two-dimensional (2D) materials, where defects in the crystalline structure drastically alter their mechanical, optical, electronic and quantum properties. Typically, such defects are characterized by means of scanning confocal microscopy[1], near-field scanning optical microscopy (NSOM)[2,3], and atomic force (AFM) or scanning tunneling (STM) microscopies[4,5]. For the atomically-high resolution of structural defects, transmission electron microscope (TEM) is often utilized[6,7], however this approach can also induce additional defects in the material under study[8]. The main disadvantage of all these characterization approaches is a small field-of-view (FOV) of up to 100 nm$^2$, which makes them incompatible with high-throughput wafer-scale processes, crucial for microelectronics and photonics industries.

On the contrary, widefield optical microscopy is routinely used for material characterization purposes, allowing for high-speed inspection, but lacking the nanoscale resolution needed to detect single defects. This disadvantage can be often compensated through the use of various super-resolution microscopy techniques, including single-molecule localization microscopy (SMLM)[9,10]. SMLM has been recently used for the characterization of optically-active defects in 2D materials[11–13], in particular those with "blinking" defects (stochastically switching between their ON and OFF states), such as in hexagonal Boron Nitride (hBN). Furthermore, the use of specialized on-chip imaging platforms[14] has drastically increased the throughput of the technique[15].

Defects, however, are not only intrinsic properties of 2D materials, occurring during their growth. Inspection and incorporation of 2D materials into functional or test devices very often requires a transfer step to move the grown 2D material onto the substrate of interest. Naturally, this process might induce additional defects[16] and, in most cases, contamination from the transfer process[17]. During the transfer, the 2D material can also crack, acquire wrinkles and trap tiny air bubbles, all of which unpredictably changes the local stress and influences the optoelectronic properties of its defects[16–19].

Growing the 2D material of interest directly on target chips can circumvent all the above-mentioned issues. While seeming straightforward, this solution is far from easy to implement, as the growth of 2D materials is typically performed on catalytic metal substrates (Cu, Fe, Pt)[20–22], which facilitate the growth process, but quench the fluorescence needed for the optical inspection. Contrary to that, microelectronics and photonics industries mostly rely on the use of semiconductors and dielectrics for device fabrication – the complementary metal oxide semiconductor (CMOS) process[23,24], so there is an ongoing challenge to develop the CMOS-compatible methods of growing 2D materials to integrate them into the existing industrial-scale fabrication processes[25].

To tackle this challenge, we report here the direct growth of few-layer hexagonal boron nitride (hBN) on silicon nitride-based waveguide imaging chips and demonstrate high-throughput widefield characterization of the as-grown material and its optically-active defects. We also compare this approach to the previously demonstrated polymer-assisted transfer process, and show the clear advantage of having an intact material, enabled by the direct growth. We further characterize the obtained 2D material using TEM, electron energy loss spectroscopy (EELS), and Raman spectroscopy, confirming the chemical and structural composition of CVD-grown boron nitride.

## Results and discussion

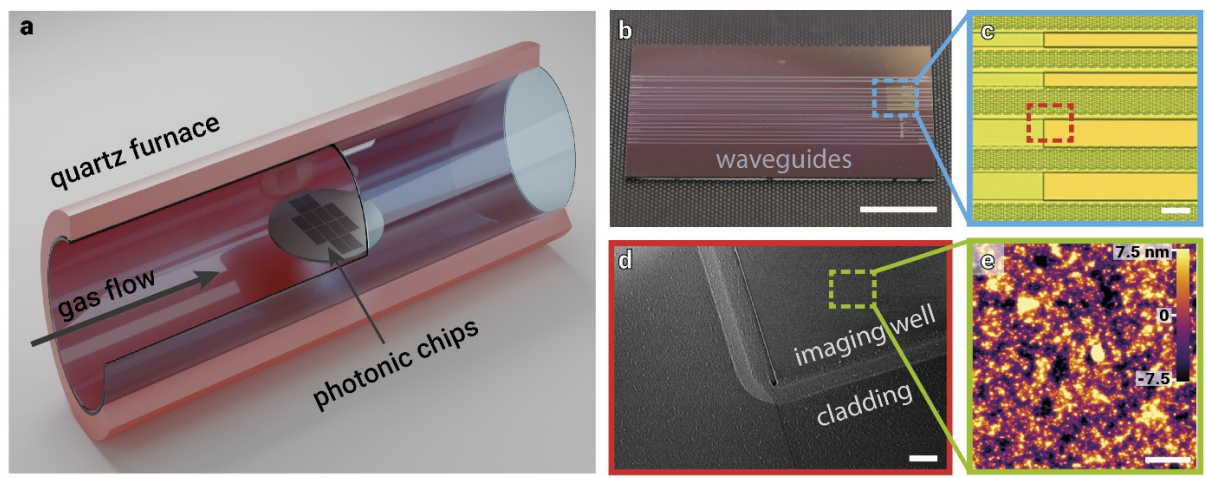

*Figure 1* Direct growth of hBN on silicon nitride photonic chips. **a)** Furnace for CVD growth of few-layer hBN with loaded silicon nitride substrates. **b)** Single 10x20 mm chip with multiple silicon nitride waveguides, running horizontally across the whole chip. Scale bar: 5 mm **c)** Optical micrograph of the zoomed-in region from b), showing a number of imaging wells on waveguides. Scale bar: 100 µm **d)** Scanning electron microscope (SEM) micrograph of a zoomed-in area next to the imaging well (dashed rectangle in c) after the growth. Scale bar: 20 µm **e)** AFM image of the directly grown few-layer hBN film on silicon nitride. Scale bar: 2 µm

We performed a low-pressure chemical vapor deposition (LPCVD) growth of hBN directly on the fabricated silicon nitride waveguides in a quartz tube furnace using ammonia borane as a precursor. The schematic of the CVD reactor is shown in Fig. 1a and the detailed description of the growth process can be found in Materials and Methods section below. Waveguide imaging chips were fabricated on silicon wafers using silicon nitride as a core layer and silicon dioxide as top and bottom cladding. Each chip contained several waveguides (Fig. 1b) with imaging wells (Fig. 1c, d) – areas where protective top cladding was removed to have direct access to the surface of the waveguide core. After extensive cleaning, waveguide chips were loaded into the furnace and left in vacuum overnight. The growth was performed at a temperature of 1200˚C for 45 minutes, which resulted in a formation of a few-layer hBN covering the whole surface of the chips[26].

While it is not straightforward to observe the uniformly-grown hBN layer on the optical micrographs (Fig. 1b,c), the SEM image (Fig. 1d) shows an increased surface roughness after the growth, which is further confirmed by the AFM analysis in Fig. 1e and Figs. S1-2. Moreover, the grown material has a different morphology inside and outside the imaging well, i.e. on crystalline silicon nitride versus amorphous silicon dioxide, which is shown in Fig. S1 and is probably rooted in the initially different surface roughness of the corresponding areas, shown in Fig. S2. Further analysis of the as-grown hBN, including Raman and Energy Dispersive X-ray spectroscopy (EDX), TEM and Electron Energy Loss Spectroscopy (EELS), is shown in the supporting information (Figs. S3-S6).

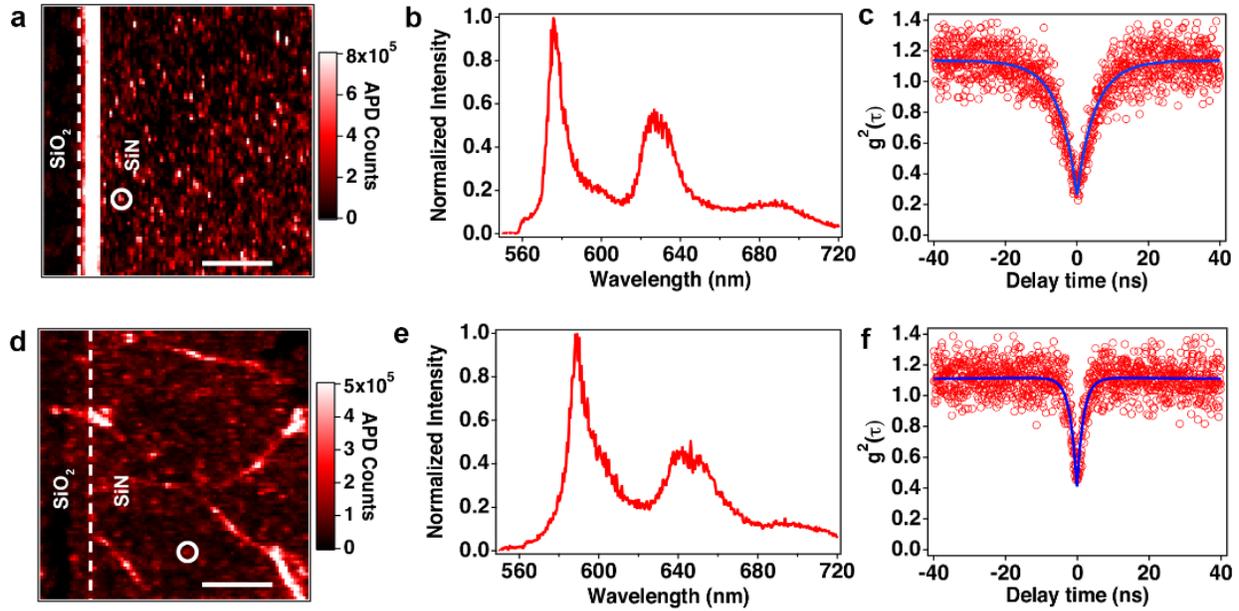

*Figure 2 Confocal imaging of directly-grown (a-c) vs transferred (d-f) hexagonal boron nitride. **a,d)** Confocal scans of the area next to the imaging well. **b,e)** Representative spectra of one of the inspected emitters from the confocal scans. **c,f)** Autocorrelation functions for the same emitters, confirming their single-photon nature. The bright lines visible in d) are supposedly transfer-induced wrinkles, possibly containing polymer residues. All measurements were done in air.*

Directly after growth, the imaging chips were inspected using a home-built confocal microscope, revealing multiple single-photon emitters (SPEs) in the as-grown material with characteristics similar to the ones reported in literature[27]. The results of this characterization are shown in Fig. 2a-c. In parallel, hBN grown via a similar CVD-process on copper foil[27] was transferred onto one of the clean imaging chips using a standard polymer-assisted transfer process[17]. This chip with transferred hBN was inspected as well using the same confocal microscope and the results of its inspection are shown in Fig. 2d-f.

Even though on both chips it is possible to identify the characteristic SPEs in hBN, when comparing the two approaches, one can easily notice the presence of cracks, folds and polymer residues in the transferred hBN and their absence in the as-grown material. Importantly, this has far-reaching consequences for large-area statistical based characterization of the optically-active defects in the material, which we further demonstrate using widefield localization microscopy.

Most of the widefield imaging presented further was performed using the custom-built waveguide microscope, schematically shown in Fig. 3a and described in detail in Ref. 14. Briefly, the imaging chips with hBN on top are placed into a custom-made sample holder and the laser light is coupled into the on-chip waveguides using a long working distance objective. The evanescent field of the propagating light excites the optically-active hBN defects on the waveguide surface (inside the imaging wells where the protective cladding has been removed). The excited region of the waveguide is highlighted in Fig. 3b by the red rectangle. The fluorescent signal is captured by a separate water-dipping objective and recorded using an sCMOS camera (see Materials and methods for the details of the imaging setup).

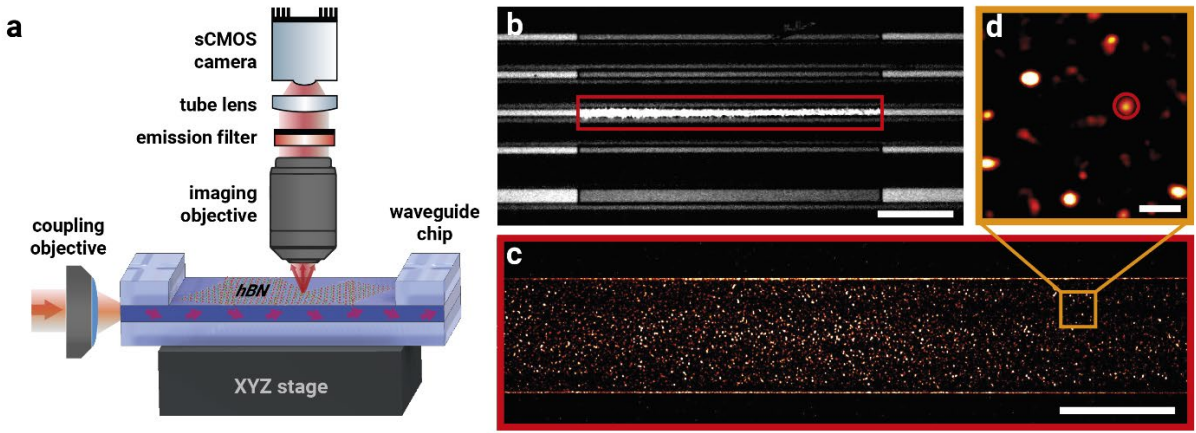

***Figure 3*** *Waveguide-based widefield imaging of optically-active emitters in as-grown hBN. **a)** Simplified schematic of the waveguide microscope. **b)** Overview of the imaging wells when the laser light is coupled into one of the waveguides (red rectangle). Scale bar: 250 µm **c)** Reconstructed image of emitters in hBN inside the imaging well, highlighted in b). Scale bar: 25 µm **d)** Zoom-in on a region in c), showing individual emitters. Scale bar: 1 µm*

Each recording consists of several thousand frames, acquired with a typical exposure time between 20 and 100 ms, which are then processed to localize the point emitters and reconstruct the super-resolved image. The result of this process is shown in Fig. 3c and one can see a relatively uniform coverage of the surface of hBN in the imaging wells by the optically-active defects. A zoom-in on one of the areas in Fig. 3d shows isolated optically-active emitters, with a typical localization uncertainty around 60 nm (Fig. 4a). This value is estimated by analysing the distribution of localizations for each emitter, shown in the inset of Fig. 4a.

To further assess the ensemble properties of the optically-active defects, we utilized a spectral super-resolution (sSMLM) technique, providing a combined spatial and spectral information about the localized defects[12]. Such capabilities could be easily integrated into the waveguide microscope by introducing a dispersive prism into the collection path, however here the existing setup with the direct excitation was used. This approach allowed us to see the spectral distribution of the optically-active emitters over the whole FOV, revealing two sets of defects ("green" and "red") activated either by 561 or 647 nm laser illumination, respectively (Fig. 4b). These two sets of emitters were previously reported in Ref. 12, where the emission peak around 580 nm is attributed to the zero-phonon line (ZPL) of the most commonly observed defect type[28]. The second peak from Fig. 4b around 610 nm is most probably originating from the phonon sideband of the same emitter type and can also be seen on spectra from Fig. 2b,e. It is important to note here, that the sSMLM technique shows a statistical distribution of spectra across the FOV, so it should be viewed as an averaged overlay of all recorded single-emitter spectra.

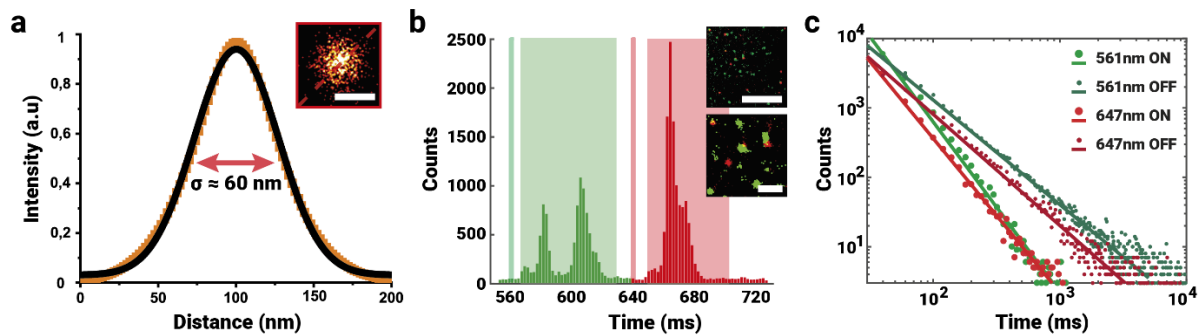

*Figure 4* Ensemble properties of optically-active emitters in as-grown hBN. *a)* Typical Gaussian intensity profile of an individual defect, highlighted emitter in Fig. 3d. The inset shows a further zoom-in into the reconstructed image of the localized emitter. Scale bar: 100 nm *b)* Spectral statistics of emitters in the as-grown hBN, obtained using an sSMLM technique. Semi-transparent lines show the laser excitation wavelengths. Coloured semi-transparent regions represent the transmission bands of the emission filters used. Insets show the spatial map of two types of emitters. Scale bars: 5 µm, 500 nm *c)* Temporal statistics of blinking emitters, showing the distribution of their ON and OFF times for two excitation wavelengths (561 nm and 647 nm).

The third peak in Fig. 4b, centered around 670 nm, most probably corresponds to an hBN defect of another type. Its emission is not very pronounced when excited with a green laser (532 or 561 nm), but is significantly enhanced upon excitation with the 647 nm laser. This defect type has been previously reported in Refs. 12,29 and is commonly seen in both exfoliated and CVD-grown hBN. The spatial distribution of both types of defects is shown in the inset of Fig. 4b, illustrating the importance of using several laser lines or a tunable light source to characterize the variety of optically-active defects in 2D materials. Finally, we also checked the ON-OFF statistics of defects in as-grown hBN (Fig. 4c), which has a similar power-law distribution for both excitation wavelengths ($\alpha_{ON} \approx -2.0$, $\alpha_{OFF} \approx -1.5$) to the ones reported in literature[12,15].

After verifying the presence and characteristics of optically-active defects in hBN, grown on top of the imaging chips, we transition to the comparison with a standard hBN film, transferred onto the imaging chip using the wet transfer method (the film was produced using the same CVD growth method on a conventional copper substrate). We first inspected the samples with both as-grown and transferred material using SEM, which allows to easily visualize possible residues and contamination. Indeed, in Fig. 5a most of the contrast on the image comes from the residual polymer film, which is further confirmed by AFM imaging in Fig. S7. Additionally, one can see multiple transfer-induced tears in the transferred hBN film. In contrast to that, the surface of the chip with an as-grown hBN looks relatively clean (Fig. 5c). A closer comparison of the surfaces of two chips is provided by AFM measurements in Figs. S1,S7.

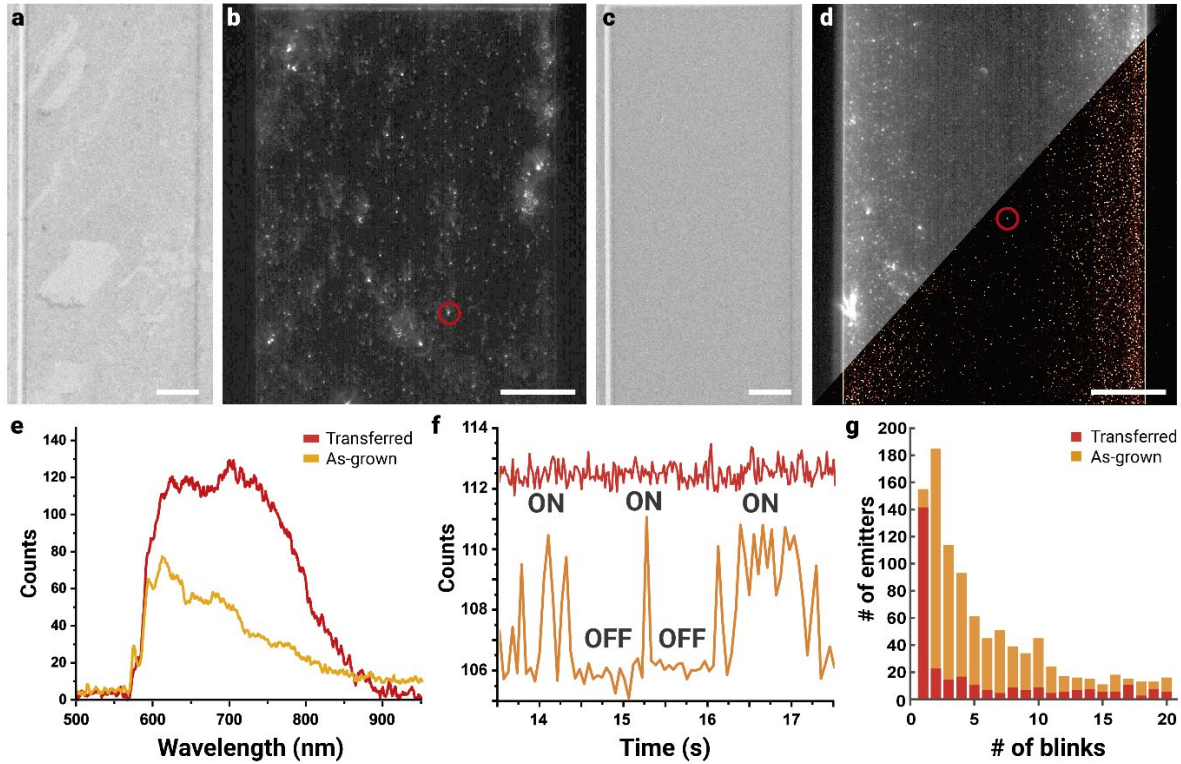

*Figure 5* Large-FOV imaging of transferred vs directly-grown hBN. *a)* SEM micrograph of transferred hBN film inside the imaging well on the waveguide chip. Residues from the wet transfer can be seen as darker regions on top of the hBN film with transfer-induced tears. *b)* Averaged stack of fluorescence images of the area inside the imaging well, obtained using the waveguide-based platform. *c)* SEM micrograph of the imaging well on the waveguide chip with directly-grown hBN. *d)* Composite image of the defects in directly-grown hBN. Imaging performed the same way as in b). Top left part shows the averaged stack of fluorescence data, while the bottom right shows the reconstructed super-resolved image. No isolated flakes are seen as hBN uniformly covers the area of the chip. Scale bar (a-d): 25 µm *e)* Typical emission spectra of the transferred vs as-grown material, obtained from the highlighted emitters in b) and d), respectively. The broad spectrum of the transferred material possibly originates from the transfer-induced contamination. *f)* Temporal intensity traces of the same highlighted emitters. *g)* Whole-FOV temporal statistics of the emitters in b) and d), confirming at the ensemble level the observed temporal dynamics seen in f). All imaging was done in DI water.

The fluorescence images of the transferred and directly grown hBN further confirm the differences in terms of sample cleanliness between the two approaches. Looking at the imaging wells of waveguide chips in Figs. 5b,d, one can see the transfer-induced imperfections and contamination in Fig. 5b. The strong background, created by them, dominates over the dimmer optically-active defects in hBN, hindering their proper localization by the SMLM algorithm. This cannot but skew the widefield characterization of optically-active defects in the inspected material.

To elaborate on that, we studied the spectral (Fig. 5e) and temporal (Fig. 5f,g) characteristics of the as-grown and transferred materials. Typical emission spectra, found on both types of samples are shown in Fig. 5e, where the as-grown material shows emission peaks characteristic of hBN, while the transferred material shows much broader spectrum which we attribute to the transfer-induced contamination. The spectral measurements were done using a confocal setup (see Materials and methods for more information).

Further comparing the temporal behaviour of emitters in both types of samples, we found that the transfer-induced residues were mostly not showing the blinking behaviour, which we typically observed for defects in hBN immersed in an aqueous solution (Fig. 5f). This is the case not just for selected emitters, but stays valid for the majority of them in more than a 100x100 µm$^2$ FOV (Fig. 5g). That is why we did not display the reconstructed image in Fig. 5b, as the non-blinking character of emitters and highly uneven background would not allow for the resolution enhancement and would create processing artefacts. On the contrary, blinking emitters in the as-grown sample allowed for successful localization and reconstruction of the super-resolved image (Fig. 5d).

The performed analysis shows how beneficial widefield SMLM technique is when inspecting 2D materials to analyse their defects at an ensemble level. This is especially true for blinking defects, which can easily go undetected during confocal scanning. Both techniques, however, are complementary to each other and help to properly validate the observed emission patterns. Their co-integration in a single optical setup, together with the capabilities of a large-FOV spectral SMLM, will be a necessary next step to increase the throughput of such optical inspection.

Naturally, 2D materials on dielectric substrates are still far from being defect-free monolayers, but even in the current state they are already useful for certain applications in nanophotonics. Constant progress in this field renders a CMOS-compatible wafer-scale process reachable in the nearest future. Such process will allow having several dedicated imaging chips per wafer, to characterize the as-grown material, while filling the rest of the wafer with functional devices. Being already a standard practice in microelectronics, integrated photonics and superconducting circuits, it is likely to be adopted by the field of 2D materials as well.

In-situ fast and high-throughput characterization of quantum emitters on waveguides can also accelerate the integration of 2D materials with nanophotonic circuits. To this extent, emitters of choice can be coupled onto the selected on-chip waveguides and addressed through them. This has already been shown with the exfoliated flakes[30], and the possibility of directly growing the 2D materials will largely simplify this process, allowing its CMOS-compatible scale-up.

## Conclusions

We demonstrated direct growth of a few-layer hBN on silicon nitride photonic chips. The embedded optically-active defects in the as-grown hBN were characterised using both a standard confocal microscope and a novel large-FOV waveguide-based imaging platform. The latter clearly unveils the advantages of the direct growth approach over wet transfer for the integration of hBN with photonic chips. The presented growth method could be extended to wafer-scale devices, diminishing the need for wet transfer of 2D materials that often results in cracks, wrinkles and polymer residues. Our approach paves the way for the future use of CVD-grown hBN in integrated photonics, coupled with high-throughput optical methods for 2D materials characterization.

## Materials and methods

### LPCVD growth of hBN

hBN was grown via low pressure chemical vapor deposition (LPCVD) in a tube furnace, using a similar protocol and setup and described in detail elsewhere[27]. Ammonia borane (Sigma Aldrich) was used as a precursor for growth. Growth was performed for 1 hour at 1200°C, a pressure of 2 torr, in an Ar/H2 (5%) environment, and the precursor was heated to 95°C for sublimation. The waveguide substrates were heated to 1200°C in an Ar/H2 environment and kept at this temperature for 1 hour prior to growth to remove any residual surface contaminants.

### Raman characterization

Raman analysis was performed on an In-Via confocal Raman (Renishaw) system, with a 633 nm excitation laser. Spectrometer calibration was performed using a blank Si substrate to 520 cm$^{-1}$.

### Confocal characterization

PL experiments were performed with a lab-built scanning confocal microscope utilizing a continuous wave (CW) 532-nm laser excitation source (Gem 532, Laser Quantum Ltd.). The CW pump laser was passed through a 532 nm line filter and a half-waveplate before focusing on the sample with a high numerical aperture objective (100×, NA =0.9, Nikon). A fast steering X–Y steering mirror (FSM-300) was used for scanning. Light emission was filtered through a 532-nm dichroic mirror (532 nm laser BrightLine, Semrock), with an additional 568-nm long pass filter (Semrock) to fully remove laser light. The sample emission was then collected through a graded-index multimode fiber with an aperture of 62.5 μm, and subsequently directed to a spectrometer (Acton Spectra Pro, Princeton Instrument Inc.) or to two avalanche photodiodes (Excelitas Technologies) in a Hanbury Brown-Twiss configuration for spectroscopy and photon counting measurements, respectively. Time-correlated single photon counting was performed via a (PicoHarp 300, PicoQuant). All g (2) (τ) measurements were analyzed and fit without background correction, and without additional spectral filtering.

### TEM characterization

Samples were imaged and characterized by the means of transmission electron microscopy (TEM). Images were obtained with 80 kV of accelerating voltage to not induce additional defects into the material. Fig. S5 was obtained using FEI Talos in the high-resolution TEM (HRTEM) mode. Fig. S6 was obtained in the scanning TEM (STEM) mode of a double-C$_s$-corrected and monochromated FEI Titan Themis in order to perform the electron energy loss spectroscopy (EELS) analysis to measure the bandgap energy and confirm the material composition.

### Widefield fluorescence imaging

Imaging was performed using a waveguide microscope described in Refs. 14,15. Briefly, laser light was coupled into the on-chip waveguide using a long-working distance objective (Mitutoyo, 50x, NA 0.55) with a nominal laser power from 50 to 100mW. The excited fluorescence was collected through a dipping objective (CFI Plan 100XC W, 100x magnification, NA 1.1) in DI water and imaged on the sCMOS camera (PRIME 95B, Photometrics). The acquisition time of the sCMOS camera was adjusted to maximize the signal-to-noise without creating overlapping localizations (typically, 50-100 ms). The acquired stacks of thousands of frames were recorded using Micro-Manager[31] and saved as TIFF files.

### Super-resolution image processing

The acquired image stacks were imported to the ImageJ software[32] and analyzed using the Thunderstorm plugin[33]. The analysis consisted of applying a wavelet filter to each frame and fitting the peak intensities by 2D Gaussian profiles. Only emitters with intensities larger than the standard deviation were then assembled into a localization table, which was used to render the super-resolved image. The temporal and spectral statistics were analysed in MATLAB.

### AFM & SEM analysis

The devices were imaged using an atomic force microscope (Asylum Research Cypher) operating in AC mode. The SEM imaging was performed using a ZEISS Leo electron microscope at 1 kV acceleration voltage. The EDX analysis was done using a ZEISS Merlin electron microscope.

## Acknowledgements

We would like to thank LIGENTEC SA and the EPFL Center of Micronanotechnology (CMi) for their help with chip fabrication and the EPFL Laboratory of Quantum Nano-Optics for the assistance with the spectral measurements. We also thank Prof. Suliana Manley for supporting our work and a generous access to the microscopy set-up. E.G. acknowledges the support from the Swiss National Science Foundation through the National Centre of Competence in Research Bio-Inspired Materials. I.A. acknowledges the The Australian Research council (CE200100010) and the Asian Office of Aerospace Research 894 and Development (FA2386-20-1-4014) for the financial support.